\documentclass[final]{svjour3}
\usepackage{gensymb}
\usepackage{graphicx}
\usepackage{rotating}
\usepackage{amssymb}
\usepackage{mathptmx}
\usepackage{comment}
\usepackage{floatrow}
\usepackage[numbers]{natbib}
\makeatletter
\journalname{Journal of Low Temperature Physics}

\bibpunct{}{}{,}{s}{}{,}

\begin{document}

\newcommand{\hdblarrow}{H\makebox[0.9ex][l]{$\downdownarrows$}-}
\title{Design and Performance of the First BICEP Array Receiver}

\author{A.~Schillaci$^1$ \and P.~A.~R.~Ade$^2$ \and Z.~Ahmed$^{3,4}$ \and M.~Amiri$^5$ \and D.~Barkats$^6$ \and R.~Basu Thakur$^1$ \and C.~A.~Bischoff$^7$ \and J.~J.~Bock$^{1,8}$ \and H.~Boenish$^6$ \and E.~Bullock$^9$ \and V.~Buza$^6$ \and J.~Cheshire$^9$ \and J.~Connors$^6$ \and J.~Cornelison$^6$ \and M.~Crumrine$^9$ \and A.~Cukierman$^4$ \and M.~Dierickx$^6$ \and L.~Duband$^{10}$ \and S.~Fatigoni$^5$ \and J.~P.~Filippini$^{11,12}$ \and G.~Hall$^9$ \and M.~Halpern$^5$ \and S.~Harrison$^6$ \and S.~Henderson$^{3,4}$ \and S.~R.~Hildebrandt$^8$ \and G.~C.~Hilton$^{13}$ \and H.~Hui$^1$ \and K.~D.~Irwin$^{3,4}$ \and J.~Kang$^4$ \and K.~S.~Karkare$^{6,14}$ \and E.~Karpel$^4$ \and S.~Kefeli$^1$ \and J.~M.~Kovac$^6$ \and C.~L.~Kuo$^{3,4}$ \and K.~Lau$^9$ \and K.~G.~Megerian$^8$ \and L.~Moncelsi$^1$ \and T.~Namikawa$^{15}$ \and H.~T.~Nguyen$^8$ \and R.~O'Brient$^{8,1}$ \and S.~Palladino$^7$ \and  N.~Precup$^9$ \and T.~Prouve$^{10}$ \and C.~Pryke$^9$ \and B.~Racine$^6$ \and C.~D.~Reintsema$^{13}$ \and S.~Richter$^6$ \and B. L.~Schmitt$^6$ \and R.~Schwarz$^6$ \and C.~D.~Sheehy$^{16}$ \and A.~Soliman$^1$ \and T.~St.~Germaine$^6$ \and B.~Steinbach$^1$ \and R.~V.~Sudiwala$^2$ \and K.~L.~Thompson$^{3,4}$ \and C.~Tucker$^2$ \and A.~D.~Turner$^8$ \and C.~Umilt\`{a}$^7$ \and A.~G.~Vieregg$^{14}$ \and A.~Wandui$^1$ \and A.~C.~Weber$^8$ \and D.~V.~Wiebe$^5$ \and J.~Willmert$^9$ \and W.~L.~K.~Wu$^{14}$ \and E.~Yang$^4$ \and K.~W.~Yoon$^4$ \and E.~Young$^4$ \and C.~Yu$^4$ \and C.~Zhang$^1$}

\institute{$^1$Department of Physics, California Institute of Technology, Pasadena, California 91125, USA
\\$^2$School of Physics and Astronomy, Cardiff University, Cardiff, CF24 3AA, United Kingdom
\\$^3$Kavli Institute for Particle Astrophysics and Cosmology, SLAC National Accelerator Laboratory, 2575 Sand Hill Rd, Menlo Park, California 94025, USA
\\$^4$Department of Physics, Stanford University, Stanford, California 94305, USA
\\$^5$Department of Physics and Astronomy, University of British Columbia, Vancouver, British Columbia, V6T 1Z1, Canada
\\$^6$Harvard-Smithsonian Center for Astrophysics, Cambridge, Massachusetts 02138, USA
\\$^7$Department of Physics, University of Cincinnati, Cincinnati, Ohio 45221, USA
\\$^8$Jet Propulsion Laboratory, Pasadena, California 91109, USA
\\$^9$Minnesota Institute for Astrophysics, University of Minnesota, Minneapolis, 55455, USA
\\$^{10}$Service des Basses Temp\'{e}ratures, Commissariat \`{a} lEnergie Atomique, 38054 Grenoble, France
\\$^{11}$Department of Physics, University of Illinois at Urbana-Champaign, Urbana, Illinois 61801, USA
\\$^{12}$Department of Astronomy, University of Illinois at Urbana-Champaign, Urbana, Illinois 61801, USA
\\$^{13}$National Institute of Standards and Technology, Boulder, Colorado 80305, USA
\\$^{14}$Kavli Institute for Cosmological Physics, University of Chicago, Chicago, IL 60637, USA
\\$^{15}$Department of Applied Mathematics and Theoretical Physics, University of Cambridge, Wilberforce Road, Cambridge CB3 0WA, UK
\\$^{16}$Physics Department, Brookhaven National Laboratory, Upton, NY 11973
\\ \email{alex78@caltech.edu}}

\maketitle

\begin{abstract}

Branches of cosmic inflationary models, such as slow-roll inflation, predict a background of primordial gravitational waves that imprints a unique odd-parity ``B-mode" pattern in the Cosmic Microwave Background (CMB) at amplitudes that are within experimental reach. The BICEP/Keck (BK) experiment targets this primordial signature, the amplitude of which is parameterized by the tensor-to-scalar ratio \emph{r}, by observing the polarized microwave sky through the exceptionally clean and stable atmosphere at the South Pole.  B-mode measurements require an instrument with exquisite sensitivity, tight control of systematics, and wide frequency coverage to disentangle the primordial signal from the Galactic foregrounds.

BICEP Array represents the most recent stage of the BK program, and comprises four BICEP3-class receivers observing at 30/40, 95, 150 and 220/270\,GHz. The 30/40\,GHz receiver will be deployed at the South Pole during the 2019/2020 austral summer. After 3 full years of observations with 30,000+ detectors, BICEP Array will measure primordial gravitational waves to a precision $\sigma$(r) between 0.002 and 0.004, depending on foreground complexity and the degree of lensing removal.  In this paper we give an overview of the instrument, highlighting the design features in terms of cryogenics, magnetic shielding, detectors and readout architecture as well as reporting on the integration and tests that are ongoing with the first receiver at 30/40\,GHz.


\end{abstract}

\section{Introduction}

Measurements of the polarization of the Cosmic Microwave Background provide key information to further our understanding of the early universe. The $\Lambda$CDM model predicts an $E$-mode polarization pattern in the CMB at the level of a few $\mu$K as well as $B$-mode polarization on arc-minute scales, arising from gravitational lensing of $E$-mode polarization by large-scale structure [1][2]. Inflationary gravitational waves source degree-scale $B$-mode polarization at an undetermined amplitude, parameterized by the tensor-scalar ratio $r$.  Measurements of inflationary B-mode polarization constrain $r$ and the energy scale of inflation.  While classes of inflation models predict undetectable low levels of $r$, a detection of inflationary $B$-mode polarization would represent strong evidence for inflation. Producing better constraints on inflationary polarization now requires subtracting polarized Galactic dust and synchrotron foregrounds and removing gravitational lensing polarization, using next-generation instruments with higher sensitivity that observe in multiple frequencies.

\begin{figure}[htbp]
\begin{center}
\includegraphics[width=1\linewidth, keepaspectratio]{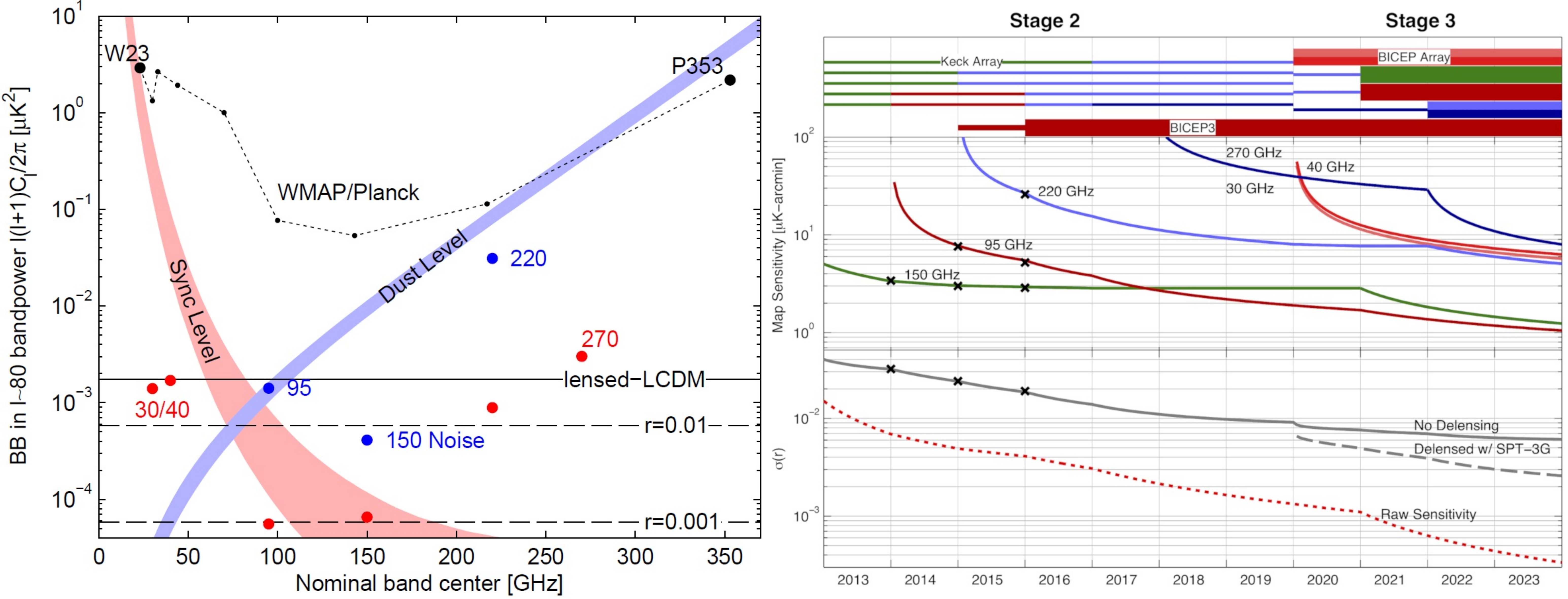}
\caption{\emph{Left:}: Expectation values and noise uncertainties for the B-mode power spectrum at degree angular scales in the BICEP/Keck field. The existing data already tightly constrains the SEDs of dust (blue band) and synchrotron (red band). Also shown are the levels of the lensing B-mode and potential inflationary signals (black lines). These can be compared to the noise uncertainties of WMAP/Planck and BICEP/Keck shown as points for the various observing frequencies. The blue points show the BK15 data set and the red ones show the projected noise levels for BICEP Array. \emph{Right:} Projected sensitivity to \emph{r} from 2013 to 2023. Top: Frequency distribution over the Stage 2 and Stage 3 receivers. Middle: Time evolution of the map depths at each frequency. Bottom: Sensitivity to \emph{r} after marginalizing over seven foreground parameters, as well as the ”no foreground” raw sensitivity (red). Crosses represent achieved sensitivities on published papers. The dashed gray line shows the expected improvement if we delens the field using SPT-3G data [Wu et al., in prep.]. (color figure online).}

\label{fig1}
\end{center}
\end{figure}

In the last decade the Bicep/Keck Collaboration [3] has pushed down limits on \emph{r} with a series of telescopes with increasing sensitivity, deployed at the South Pole Station. 
The BICEP/Keck instruments were designed specifically to control systematic errors on degree-scale measurements (multipoles 35 $<$ \emph{l} $<$ 300) using cryogenic wide-field refracting telescopes combined with arrays of planar antenna-coupled polarization-sensitive transition edge superconducting (TES) bolometers. 
To accurately measure polarized foregrounds, Keck Array carried out multi-frequency observations with 5 telescopes, accumulating years of data at 95, 150, 220 and at 270\,GHz. 
Each Keck telescope was equipped with 4 detector tiles fabricated from 100\,mm wafers, accommodating sub-arrays of 8x8 pixels (6x6 at 95\,GHz), each pixel operating two TES bolometers sensing vertical and horizontal polarization, for a total of 512 detectors per receiver (288 at 95\,GHz). 
Data accumulated through 2015, including the first 220\,GHz data, yields the tightest inflationary constraints to date of $\emph{r}_{0.05}<0.07$ at $95\%$ confidence with $\sigma(\emph{r})=0.02$ [3]. 
Since 2015, Keck observations combined with data from the BICEP3 [4] telescope and its 2560 TES bolometers operating at 95\,GHz, provide a significant increase in sensitivity.  A delensing analysis, combining this dataset with high angular resolution data from the South Pole Telescope, will further improve constraints on inflationary polarization [see Wu et al., in prep.].

BICEP Array (BA) is the next stage (\emph{Stage 3}) instrument for the BK Collaboration, replacing the Keck Array. We plan to deploy the first 30/40\,GHz BA receiver during the 2019-2020 austral summer, while keeping 3 Keck Array receivers on the sky with the new mount. Subsequent BA receivers will deploy in following seasons, eventually replacing all the Keck receivers. After three years of observations with the full BA system, the projected sensitivity reaches $\sigma(\emph{r})\leq0.004$ (see Fig.~\ref{fig1}). The final sensitivity depends on the degree of foreground complexity and the efficacy in removing the B-mode signal from gravitational lensing.

\section{BICEP Array Telescope Design}

BICEP Array [5][6] comprises 4 BICEP3-class telescopes with a total count of 30000+ photon-noise limited polarization-sensitive detectors. 
In order to fully characterize the Galactic dust and synchrotron polarized foregrounds, we have chosen the following frequencies: 30/40, 95, 150 and 220/270\,GHz. Based on previous on-sky performance, the expected sensitivities for the 4 receivers are shown in Table~\ref{tab:1}.

\begin{table}  
    \centering
    \begin{tabular}{|p{2cm}||p{2cm}|p{2cm}|p{2cm}|p{2cm}|}
  \hline
  Receiver observing band (GHz) & Number of detectors & Single detector NET  (\( \mu{{\mathrm{K}}_{\mathrm{CMB}}}\sqrt{{\rm s}}\)) & Beam FWHM (arcmin) & Survey weight per year (\({\mu}{{\mathrm{K}}_{\mathrm{CMB}}}^{-2} {\mathrm{year}}^{-1} \))\\
 \hline
 Keck Array   &&&&   \\
95 &   \textbf{288}  & \textbf{288}   & \textbf{43} & \textbf{24,000} \\
150 & \textbf{512} & \textbf{313} &  \textbf{30} & \textbf{30,000} \\
220 & \textbf{512} & \textbf{837} &  \textbf{21} & \textbf{2000} \\
270 & \textbf{512} & 1310  & \textbf{17} & 800\\
 \hline
 BICEP3 &&&& \\
95 & \textbf{2,560} & \textbf{288} & \textbf{24} & \textbf{213,000} \\
 \hline
 BICEP Array &&&& \\
$/$30 & 192 & 260 & 76 & 19,500 \\
$\setminus$ 40 & 300 & 318 & 57 & 20,500 \\
95 & 4,056 & 288 & 24 & 337,400 \\
150 & 7,776 & 313 & 15 & 453,000 \\
$/$220 & 8,112 & 837 & 11 & 32,000 \\
$\setminus$ 270 & 13,068 & 1,310 & 9 & 21,000 \\

 \hline
 
\end{tabular}
    \caption{Receiver parameters as used in sensitivity projections. Boldface numbers are actual/achieved quantities for existing receivers on the sky. The remaining values in the survey weight column are scaled from achieved survey weights using only the ratio of the number of detectors, plus, if necessary to change frequency, the ratio of nominal NET values squared. The on-sky performance of the Keck 270GHz receiver is still being evaluated.}

    \label{tab:1}
\end{table}

\begin{figure}[htbp]
\begin{center}
\includegraphics[width=1\linewidth, keepaspectratio]{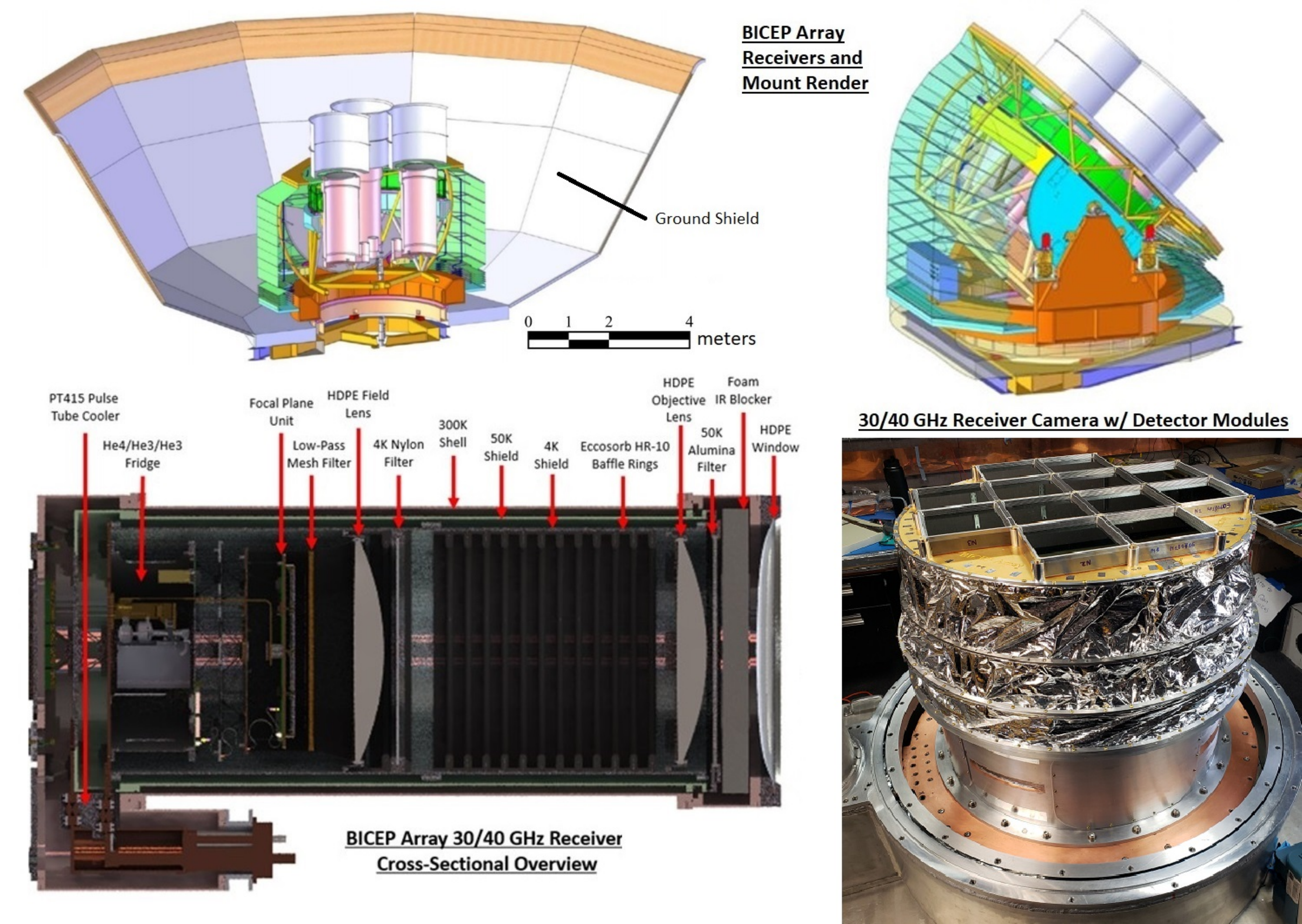}
\caption{\emph{Top/Left:} BICEP Array mount in its ground shield. \emph{Top/Right:} Detailed view of the mount with the 4 receivers in place.  \emph{Bottom/Left:} Receiver section view.  \emph{Bottom/Right:} Sub-Kelvin Insert of 30/40\,GHz BICEP Array receiver with the full complement of 12 Detector Modules installed (color figure online).}

\label{fig2}
\end{center}
\end{figure}

\begin{figure}[htbp]
\begin{center}
\includegraphics[width=1\linewidth, keepaspectratio]{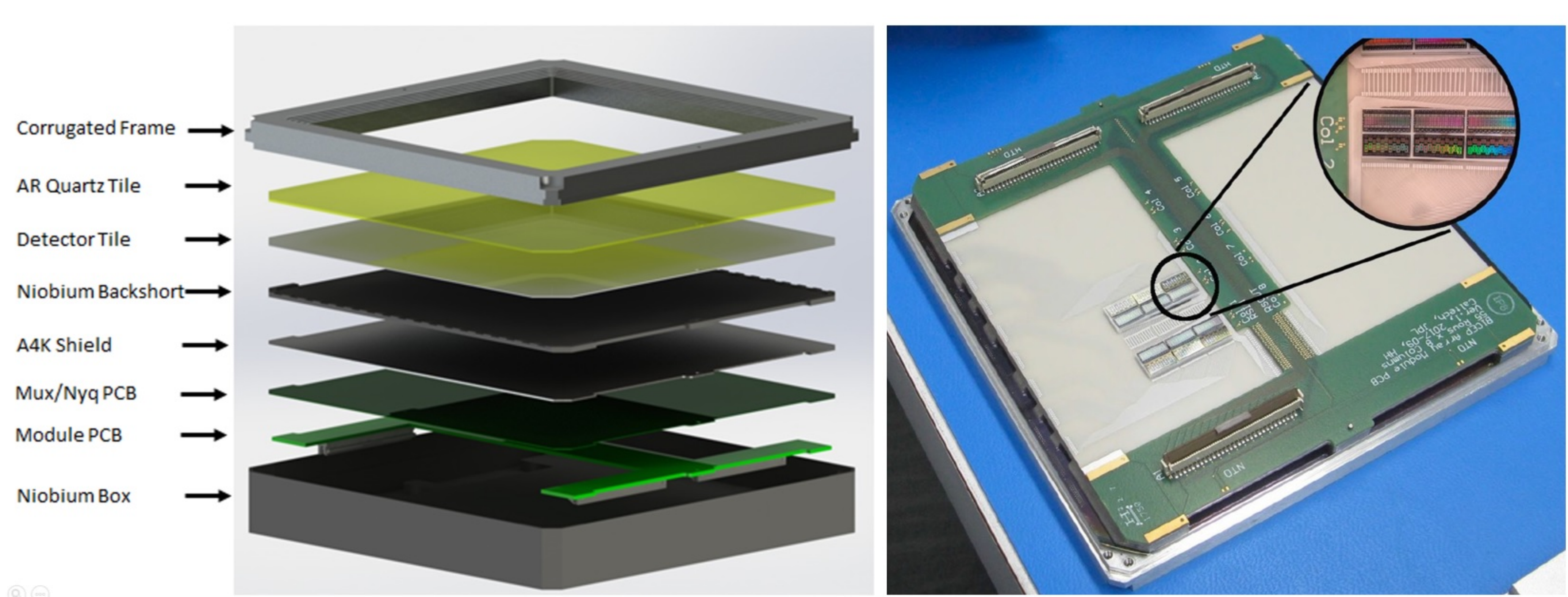}
\caption{\emph{Left:} Design of a focal plane module that houses the detector wafer, anti-reflection wafer, backshort, SQUID readout, and integrated magnetic shielding, shown in an exploded 3D rendering. \emph{Right:} An open module showing SQUID chips mounted on a printed wiring board on the back of the detector backshort (color figure online).}

\label{fig3}
\end{center}
\end{figure}

The 30/40\,GHz receiver (see Fig.~\ref{fig2}) uses a Cryomech PT415 pulse tube coupled with a He4/He3/He3 sorption fridge to cool the focal plane down to 250\,mK. The Pulse Tube is driven by a stepper motor that is physically detached and electrically isolated from the Cold Head. Residual mechanical vibrations induced by the fittings connecting the head and the remote motor are damped by a spring bellows connecting the cold head to the cryostat shell. We also use heat straps with flexible sections of braided copper wire in order to minimize the rigidity and the transmission of vibration to the cryostat stages. Each focal plane will be populated with 12 detector modules (see Fig.~\ref{fig3}) connected to the readout electronics using superconducting Nb-Ti cryogenic cables, which are carefully heat-sunk at the different temperature stages in order to minimize the conducted heat load. The cryogenic telescope is a f/1.55 refractor with a 550\,mm clear aperture that uses two alumina lenses\footnote{The 1st receiver will be deployed with HDPE lenses due to acceptable absorption loss at 30/40\,GHz.} to achieve an illuminated 475\,mm wide flat focal plane. An anti-reflection coating is applied on both lens surfaces to minimize in-band reflections. The optical filtering follows the successful BICEP3 scheme [4]. Starting from the 300\,K HDPE window, we use a 12-layer stacked HD30 Zotefoam filter (each layer 1/8" thick with 1/10" spacers) to block thermal radiation from 300\,K. Absorptive filters at 50\,K and 4\,K lower the radiative heat load on the sub-K stages, the former made of alumina and the latter made of Nylon. Finally a low-pass metal-mesh filter at 300\,mK defines the upper edge of the band.

\begin{figure}[htbp]
\begin{center}
\includegraphics[width=1\linewidth, keepaspectratio]{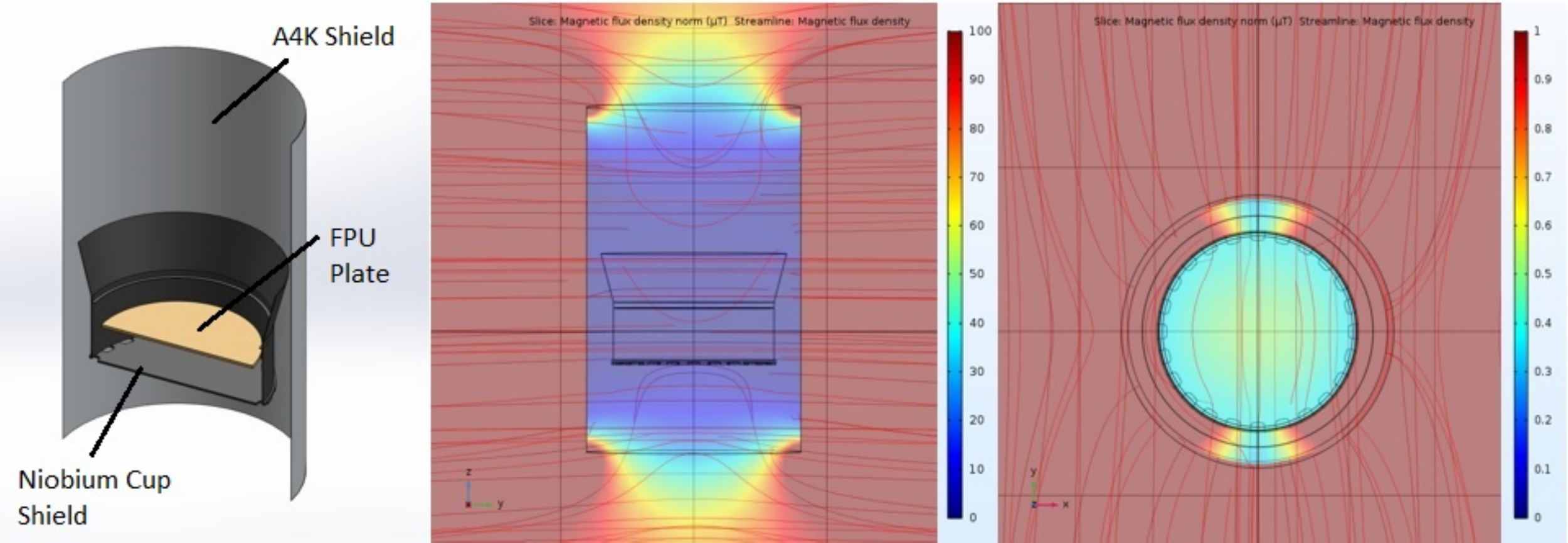}
\caption{\emph{Left:} BICEP Array receiver magnetic shields showing the A4K cylinder (light grey), the superconducting Niobium flared cup (dark grey) and the FPU plate (gold). \emph{Center:} Comsol Simulation of the Transverse residual field (vertical section, external field is $100\, \mu$T). \emph{Right} Comsol Simulation of the Transverse residual field (horizontal section at the FPU, external field is $100\, \mu$T). Color bar units are in $\mu$T. The modules provide an additional level of shielding at the focal plane (color figure online).}

\label{fig4}
\end{center}
\end{figure}

The first 3 receivers will be equipped with state-of-the-art dual-polarization antenna-coupled TES arrays, fabricated on 150\,mm diameter silicon wafers. The detector arrays are based on planar in-phase combined slot antennas that provide higher pixel packing density than feedhorns, given a chosen edge taper. The readout architecture is based on Time-Domain Multiplexing (TDM), developed by the University of British Columbia and NIST, with two stages of SQUIDs driven using a Multi-Channel Electronics (MCE) system that controls SQUID addressing and digitizes the detector signals. All the readout chain is based on numerous previous successful experiments [e.g., 4], with the detector modules carrying the first-stage SQUIDs on a PCB located behind the $\lambda$/4 backshort (see figure \ref{fig3}) and connected through a custom distribution board and superconductive NbTi wires to a stage of Series Array SQUIDs at 4K. In the design of these PCBs and the architecture of the cables, we paid particular attention to pairing signal and return in order to minimize cross-talk among channels. Moreover, SQUIDs and TES detectors must be carefully shielded from sources of magnetic fields [7][8], including the Earth's field. We studied multiple configurations using COMSOL Multi-Physics software (see figure \ref{fig4})[9]. We selected an architecture with a cylindrical A4K high magnetic permeability shield at 50\,K, combined with a Niobium superconductive flared cup at 300\,mK, which provides a shielding factor of 200 at the FPU. The TESs and the SQUIDs are then shielded inside the focal plane module, using a Niobium superconducting case and an interior layer of A4K to provide an additional shielding factor of about 500. We estimate that the full system provides shielding a factor between 50,000 to 100,000 at the SQUIDs.

We will perform direct measurements of the system to validate the magnetic field simulations by applying an external field that is modulated at a known frequency.  We currently have installed two modules where the detector and squids are replaced with flux gate sensors capable of very sensitive magnetic field measurements.

The first 30/40\,GHz receiver (BA1) will be populated with 6 modules at each frequency in a checkerboard arrangement. The design for the two frequencies is similar, with appropriate dimensions of the antenna and module scaled accordingly to frequency. We designed the module frame wall with broad-band corrugations to minimize polarized beam mismatch on the edge pixels in both frequency bands [10].

The design of the following two receiver cameras at 95\,GHz (BA3) and 150\,GHz (BA2) is at an advanced level of development and it is based on the same Time Multiplexing architecture and TES detector arrays. We are currently manufacturing prototype high-density module PCBs that allow an 18x18 pixel 150\,GHz tile that will result in a 7776 detector receiver. As the last 220/270\,GHz receiver (BA4) has over 20,000 detectors, we are pushing on RF-multiplexing to reduce the complexity of the readout system. One possible implementation is the microwave SQUID readout ($\mu$Mux) system that has been already been operated with our current TES detectors for a year as an on-sky demonstrator [11]. This system uses the same TDM detectors architecture coupled with an RF interface. Alternatively, Thermal Kinetic Inductance Detectors (TKID) offer on-wafer RF multiplexing to greatly simplify wafer hybridization, and are currently being developed and tested within the collaboration. 

We have developed a new multi-receiver telescope mount to replace the Keck mount at South Pole Station. Following the Keck and BICEP3 designs, the mount provides bore-sight rotation about the optical axis, in addition to rotation in Azimuth and Elevation. We plan to install the BA1 cryostat plus 3 Keck receivers in the available 4 mount slots in the coming 2019/2020 austral summer. The observations will continue to focus on the deep BICEP/Keck sky patch so that the new low-frequency data can be combined with existing multi-frequency observations.

\section{BICEP Array 30/40\,GHz receiver performance}

The 30/40\,GHz BA1 cryostat was successfully integrated and tested and we achieved good cryogenic performance in the Summer of 2019, including the full window and thermal filters. At the same time, we commissioned a second receiver with a small array of focal plane detectors that has been fully integrated with the mount.

We have integrated and tested six 40\,GHz detector modules and recently integrated the first 30\,GHz unit. Each module is tested either in a separate cryostat or in the BA1 receiver. Each module undergoes electrical testing under dark conditions, and optical testing through the optics and window.  In a dark run we measure the TES properties, thermal isolation, and noise performance. We carefully optimized the 30 and 40\,GHz detectors for the expected optical loading at the South Pole. The 40\,GHz detectors are performing near photon-limited sensitivity, with a NET of about 360$\mu$ $K_{CMB}/\sqrt{Hz}$[12]. 

\begin{figure}[htbp]
\begin{center}
\includegraphics[width=1\linewidth, keepaspectratio]{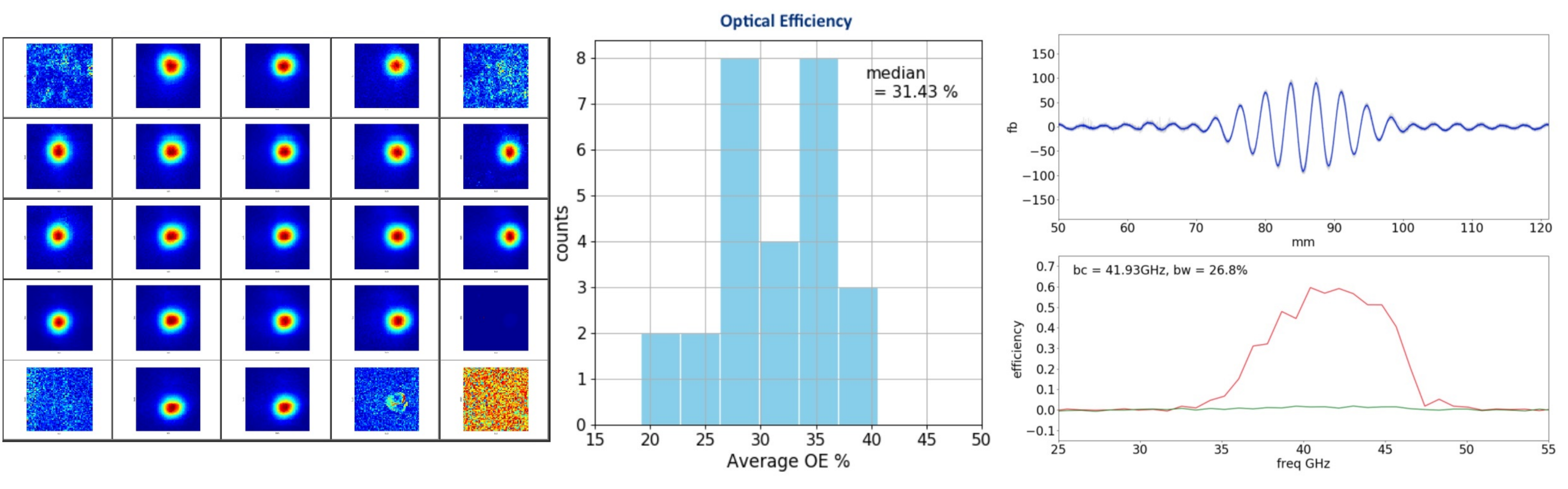}
\caption{Results from optical characterization of 40\,GHz detectors.  \emph{Left:} Antenna beam maps over a module, showing a single polarization.  \emph{Center:} A histogram of measured optical efficiency from 300 K and 77 K sources.  These measurements are end-to-end and included losses in the thermal filters and window.  \emph{Right:} Interferogram and spectral response for a typical 40\,GHz detector (color figure online).}

\label{fig5}
\end{center}
\end{figure}

In a light run we measure end-to-end optical efficiency, spectral response, and antenna beam patterns [13][14]. The detectors use a high-background aluminum TES to facilitate lab measurements, avoiding an optical attenuator that can introduce polarized artifacts.  In Fig~\ref{fig5} we show a typical beam map, optical efficiency histogram and spectral response. Tested 40\,GHz modules show low differential beam ellipticity, an average end-to-end efficiency exceeding $30\%$ and a band center (width) of 41.9\,GHz ($27\%$). These values are in line with our design targets; for a more exhaustive discussion, see [12][13][14].

\begin{acknowledgements}
The BICEP/Keck project have been made possible through a series of grants from the National Science Foundation including 0742818, 0742592, 1044978, 1110087, 1145172, 1145143, 1145248, 1639040, 1638957, 1638978, 1638970, \& 1726917 and by the Keck Foundation.The development of antenna-coupled detector technology was supported by the JPL Research and Technology Development Fund and NASA Grants 06-ARPA206-0040, 10-SAT10-0017, 12-SAT12-0031, 14-SAT14-0009 \& 16-SAT16-0002. The development and testing of focal planes were supported by the Gordon and Betty Moore Foundation at Caltech.Readout electronics were supported by a Canada Foundation for Innovation grant to UBC.The computations in this paper were run on the Odyssey cluster supported by the FAS Science Division Research Computing Group at Harvard University. The analysis effort at Stanford and SLAC is partially supported by the U.S. DoE Office of Science.We thank the staff of the U.S. Antarctic Program and in particular the South Pole Station without whose help this research would not have been possible.Tireless administrative support was provided by Kathy Deniston, Sheri Stoll, Irene Coyle, Donna Hernandez, and Dana Volponi. 
\\This is a post-peer-review, pre-copyedit version of an article published in Journal of Low Temperature Physics. The final authenticated version is available online at: http://dx.doi.org/10.1007/s10909-020-02394-6.

\end{acknowledgements}

\pagebreak

\end{document}